\documentclass[11pt]{amsart}
\usepackage{amssymb}
\newtheorem{rem}{Remark}

\newtheorem{prob}{Problem}
\newcommand{\abs}[1]{\left\vert#1\right\vert}

\newcommand{\set}[1]{\left\{#1\right\}}
\newcommand{\Real}{\mathbb{R}}
\newcommand{\Integer}{\mathbb{Z}}
\newcommand{\Complex}{\mathbb{C}}
\newcommand{\To}{\longrightarrow}

\newcommand{\liealg}[1]{\mathfrak{#1}}

\newcommand{\hilb}{\mathcal{H}}
\newcommand{\gr}[2]{\textrm{Gr}_{#1}(#2)}
\newcommand{\unit}[1]{\textrm{U}(#1)}
\newcommand{\tr}[1]{\textrm{tr}(#1)}

\newcommand{\ci}{\textrm{i}}
\begin{document}
\title{Cyclic Evolution on Grassmann Manifold and Berry Phase}
\author{Zakaria Giunashvili}%%%
\address{A. Razmadze Mathematics Institute Georgian Academy of Sciences}%
\thanks{The work is supported by Georgian National Scientific Foundation (Grant No GNSF/ST06/4-050)}%
\begin{abstract}
For a given $k$-dimensional subspace $V_0$ in a Hilbert space $\hilb$ and a
unitary transformation $g_0:V_0\To V_0$, we find a path in the Grassmann
manifold the monodromy of which coincides with $g_0$.
\end{abstract}
% ----------------------------------------------------------------
\maketitle
%%%%%%%%%%%%%%%%%%%%%%%%%%%%%%%%%%%%%%%%%%%%%%%%%%%%%%%%%%%%%%%%%%%%%%%%%%%%%%%%%%%%%%%
Let $\hilb$ be a finite-dimensional Hilbert space; $\unit{\hilb}$ be the Lie group of
unitary transformations of $\hilb$ and $\liealg{u}(\hilb)$ be the corresponding Lie
algebra. For any positive integer $k$, the Grassmann manifold $\gr{k}{\hilb}$ is defined
as the set of all $k$-dimensional subspaces of $\hilb$. This manifold can also be
dscribed as the set of corresponding orthogonal projectors
$$
\gr{k}{\hilb}=\set{P:\hilb\To\hilb\mid P\textrm{ is linear },\ P^\dag=P,\ \tr{P}=k}.
$$
As it is well-known for any given Hamiltonian $H\in\liealg{u}(\hilb)$ the corresponding
Schr\"{o}dinger equation is defined as the equation of the form
\begin{equation}\label{shred_eqn}
\dot{\psi}(t)=H(\psi(t)),\quad\psi(t)\in\hilb,\ t\in\Real,\ \psi(0)=\psi_0,
\end{equation}
and
$$
\Phi_H=\set{\exp(tH)\mid t\in\Real}
$$
is the corresponding one-parameter family of unitary transformations of $\hilb$

Obviously, the equation (\ref{shred_eqn}) defines a dynamical system on the Grassmann
manifold $\gr{k}{\hilb}$:
\begin{equation}\label{grass_shred_eqn}
\dot{P}(t)=[H,P(t)],\quad t\in\Real
\end{equation}
and the corresponding one-parameter group of diffeomorphisms of $\gr{k}{\hilb}$ is
defined by the action of the group $\Phi_H$ on $\gr{k}{\hilb}$. The action of the group
$\Phi_H$ for the projector representation of $\gr{k}{\hilb}$, is
$$
P\mapsto\exp(tH)P\exp(-tH).
$$
For a given $k$-dimensional subspace $V_0\in\hilb$, we are interested in Hamiltonians
$H\in\liealg{u}(\hilb)$ such that, after the time period $t=1$, the one-parameter group
$\Phi_H$ brings $V_0$ to itself. In other words, for a given point $P_0\in\gr{k}{\hilb}$
we are looking for Hamiltonians $H\in\liealg{u}({\hilb})$ such that the trajectory of the
equation (\ref{grass_shred_eqn}) through the point $P_0$ is closed:
$$
\exp(H)P_0\exp(-H)=P_0.
$$
When the transformation $\exp(H)$ brings the subspace $V_0$ to itself, it defines a
unitary transformation
$$
g_0=\exp(H)\vert_{V_0}:V_0\To V_0.
$$
\begin{rem}
In fact, the unitary transformation $g_0:V_0\To V_0$, induced by the one-parameter flow
$\set{\exp(tH)\mid t\in\Real}$, is the well-known Berry phase and can be decomposed in so
called ``dynamical'' and ``geometrical'' factors. Here we don't concern this
decomposition and consider the Berry phase as a ``single whole''.
\end{rem}
After this, we can reformulate our problem as
\begin{prob}\label{problem}
for a given $k$-dimensional subspace $V_0\in\hilb$ and a unitary transformation
$g_0:V_0\To V_0$, find a skew-hermitian operator $H:\hilb\To\hilb$ such that
$\exp(H)V_0=V_0$ and $\exp(H)\vert_{V_0}=g_0$.
\end{prob}
It is clear that when $[H,\ P_0]=0$, the solution of the Schr\"{o}dinger equation
(\ref{grass_shred_eqn}) with the initial condition $P(0)=P_0$ is constant: $P(t)=P_0,\ t\in[0,\ 1]$, %
therefore, it is preferable that the operator $H$ be such that $[H,\ P_0]\neq0$.
\begin{rem}
In \cite{nakahara} it is considered the similar problem, but for the ``geometric'' factor
of the Berry phase corresponding to the cyclic trajectory on the Grassmannian
$\gr{k}{\hilb}$ defined by $\exp(tH),\ t\in[0,1]$.
\end{rem}
Further we will discuss the solution of Problem \ref{problem}.

Let $m=\dim(V_0)$ and $\mathcal{E}_0=\set{e_0,\ldots,e_m}$ be an orthonormal basis of
$V_0$ consisting of eigenvectors of the operator $g_0$:
$$
g_0(e_k)=u_k\cdot e_k,\quad u_k\in\Complex,\ \abs{u_k}=1,\ k=1,\ldots,m.
$$
Consider an orthonormal extension of the basis $\mathcal{E}_0$ to the basis of the entire
Hilbert space $\hilb$:
$$
\mathcal{E}=\mathcal{E}_0\bigcap\mathcal{E}_1,\quad\mathcal{E}_1=\set{e_{m+1},\ldots,e_{n}}\subset V_0^\perp,%
$$
where $n=\dim(\hilb)$, and define the unitary operator $g:\hilb\To\hilb$ as
$$
g\vert_{_{V_0}}=g_0,\quad g(e_{m+1})=u_m\cdot e_{m+1}\quad\textrm{and}\quad g(e_p)=e_p\quad\textrm{for}\quad m+2\leq p\leq n.%
$$
In other words, we set that the vectors $e_1,\ \ldots\ ,\ e_m,\ e_{m+1},\ \ldots\ ,\ e_n$
are eigenvectors of $g$, the restriction of the operator $g$ to the subspace $V_0$
coincides with $g_0$, the eigenvalues of $g$ on $e_m$ and $e_{m+1}$ are equal and its
eigenvalues on the vectors $e_{m+2},\ldots,e_n$ are equal to 1. The matrix of the
operator $g$ in the basis $\mathcal{E}$ is of the form
$$
U=
\begin{bmatrix}
  u_1 &    0    &    &  \cdots   &   &       & 0 \\
   0   & \ddots &  \ddots   &     &   &       &   \\
      &   \ddots     & u_m &   0  &   &       &  \vdots \\
    \vdots  &        & 0    & u_m & \ddots  &       &   \\
      &        &     &  \ddots   & 1 &       &   \\
      &        &     &     &   &\ddots & 0 \\
   0   &        &   \cdots  &     &   &    0   & 1 \\
\end{bmatrix},
$$
and the matrix of the projector $P_0$ in the same basis is
$$
A=
\begin{pmatrix}
  \mathbf{1}_m & \mathbf{0} \\
  \mathbf{0} & \mathbf{0}_{n-m} \\
\end{pmatrix},
$$
where $\mathbf{1}_m$ denotes $m\times m$ identity matrix and $\mathbf{0}_{n-m}$ denotes
$(n-m)\times(n-m)$ zero matrix. Hence, the problem is reduced to the finding a matrix $H$
such that $\exp(H)=U$ and $[H,\ A]\neq0$.

Assume $u_1=e^{\ci\lambda_1},\ldots,u_m=e^{\ci\lambda_m},\ \lambda_k\in\Real,\ k=1,\ldots,m$. %
Obviously, the number $u_m$ can also be written as $u_m=e^{\ci(\lambda_m+2\pi n)},\ n\in\Integer$. %
For any unitary transformation $\omega\in\unit{2}$ let $H\equiv H_\omega$ be the
following block-diagonal matrix
$$
H_\omega=
\begin{pmatrix}
  H_1 & 0 & 0 \\
  0 & \Omega & 0 \\
  0 & 0 & \mathbf{0}_{n-m-1} \\
\end{pmatrix},
$$
where $H_1$ is the $(m-1)\times(m-1)$ diagonal matrix: $H_1=\textrm{diag}[\ \ci\lambda_1,\ \ldots,\ \ci\lambda_{m-1}\ ]$; %
and $\Omega$ is the matrix
$$
\Omega=
\omega
\begin{pmatrix}
  \ci\lambda_m & 0 \\
  0 & \ci(\lambda_m+2\pi n) \\
\end{pmatrix}
\omega^{-1}.
$$
It is clear that $\exp(H_\omega)$ is
$$
\exp(H_\omega)=
\begin{pmatrix}
  \exp(H_1) & 0 & 0 \\
  0 & \exp(\Omega) & 0 \\
  0 & 0 & \mathbf{1}_{n-m-1} \\
\end{pmatrix}.
$$
Since
$$
\exp(\Omega)=%
\omega\exp
\begin{pmatrix}
  \ci\lambda_m & 0 \\
  0 & \ci(\lambda_m+2\pi n) \\
\end{pmatrix}
\omega^{-1}
=
\begin{pmatrix}
  u_m & 0 \\
  0 & u_m \\
\end{pmatrix},
$$
we obtain $\exp(H_\omega)=U$. On the other hand, it is clear that $[H_\omega,\ A]=0$ if
and only if $[\epsilon,\ \Omega]=0$, where
$$
\epsilon=
\begin{pmatrix}
    1 & 0 \\
    0 & 0 \\
\end{pmatrix},
$$
and the latter happens only when $\Omega$ is diagonal.

To summarize, we can say that we have a family of solutions of Problem \ref{problem}
depending on the unitary matrix $\omega\in\unit{2}$ and the integer $n$.
%%%%%%%%%%%%%%%%%%%%%%%%%%%%%%%%%%%%%%
\bibliographystyle{amsplain}

\end{document}